\documentclass[aps,pre,twocolumn]{revtex4-2}

\usepackage{bm}
\usepackage{mathrsfs,amsbsy,amssymb,latexsym,amsfonts,amsmath,fancyhdr}
\usepackage[dvipdfmx]{graphicx}
\usepackage{epsfig}
\usepackage{color}
\usepackage{url}
\usepackage[normalem]{ulem}
\usepackage[hidelinks]{hyperref}
\usepackage{mhchem}

\newcommand{\pder}[2]{\frac{\partial #1}{\partial  #2}}

\newcommand{\der}[2]{\frac{d #1}{d  #2}}

\newcommand{\kB}{k_\mathrm{B}}
\newcommand{\nm}{\nonumber\\}
\newcommand{\ep}{\varepsilon}
\newcommand{\eq}{\mathrm{eq}}

\newcommand{\eff}{\mathrm{eff}}

\newcommand{\subL}{\mathrm{L}}
\newcommand{\subG}{\mathrm{G}}

\newcommand{\subC}{\mathrm{c}}
\newcommand{\subS}{\mathrm{s}}

\newcommand{\Ps}{p_\mathrm{s}}
\newcommand{\bT}{\tilde T}
\newcommand{\bP}{\bar p}

\newcommand{\Tc}{T_\mathrm{c}}

\newcommand{\xo}{x_\mathrm{o}}
\newcommand{\xb}{0}
\newcommand{\xt}{L}
\newcommand{\xs}{x_\theta}
\newcommand{\xm}{x_\mathrm{m}}
\newcommand{\Tt}{T_2}
\newcommand{\Tb}{T_1}

\newcommand{\subl}{\ell}
\newcommand{\subu}{\mathrm{u}}

\newcommand{\NleL}{N^\mathrm{L}_\mathrm{eq}}

\newcommand{\VleL}{V^\mathrm{L}_\mathrm{eq}}

\newcommand{\NlsL}{N^\mathrm{L}_\mathrm{s}}
\newcommand{\NlsG}{N^\mathrm{G}_\mathrm{s}}
\newcommand{\VlsL}{V^\mathrm{L}_\mathrm{s}}
\newcommand{\VlsG}{V^\mathrm{G}_\mathrm{s}}
\newcommand{\NL}{N^\mathrm{L}}
\newcommand{\VL}{V^\mathrm{L}}
\newcommand{\Ng}{N^\mathrm{G}}
\newcommand{\VG}{V^\mathrm{G}}
\newcommand{\Nl}{\mathcal{N}^{\ell}}
\newcommand{\Vl}{\mathcal{V}^{\ell}}

\begin{document}

\title{Thermodynamic Variational Principle Unifying Gravity and Heat Flow}

\author{Naoko Nakagawa}
\email{naoko.nakagawa.phys@vc.ibaraki.ac.jp}
\affiliation{Department of Physics, Ibaraki University, Mito 310-8512, Japan}
\author{Shin-ichi Sasa}
\email{sasa.shinichi.6n@kyoto-u.ac.jp}
\affiliation{Department of Physics, Kyoto University, Kyoto 606-8502, Japan}

\date{\today}

\begin{abstract}
Predicting the stable phase configuration in a liquid-gas system becomes a fundamental challenge when the stratification favored by gravity conflicts with arrangements induced by heat flow, particularly because standard equilibrium thermodynamics is insufficient in such non-equilibrium steady states.
We propose a variational principle based on an extended thermodynamics, called global thermodynamics, to address this state selection problem. Our key finding is that gravity and heat flow effects are unified into a single parameter, ``effective gravity'' ($g_\eff$), within this framework. Crucially, the sign of $g_\eff$ determines the stable configuration: liquid is at the bottom if $g_\eff > 0$, and floats above the gas if $g_\eff < 0$. This provides a quantitative tool for the configuration prediction under competing drives.
\end{abstract}

\maketitle

The interplay between gravity and heat flow presents fundamental questions concerning phase configurations in fluid systems.
Gravity favors stratification with denser phases at the bottom, while heat flow can induce non-equilibrium arrangements, often concentrating liquid near colder boundaries. 
When these influences conflict, for instance, with top cooling under gravity, a rich variety of phenomena are observed in experiments \cite{Zhong, Ahlers, Urban} and in nature \cite{quere2013,quere2019,dror2023}. 
Even in the linear response regime from an equilibrium state, predicting the stable liquid-gas configuration in a non-equilibrium steady state (NESS) is non-trivial, posing a challenge that lies beyond local equilibrium descriptions \cite{Landauer88}.

While a deterministic hydrodynamic description is a standard approach \cite{Landau-Lifshitz-Fluid}, its application to phase-coexisting systems is complicated by the singularly thin nature of the interface. 
This can lead to the existence of multiple stationary solutions, trapping the deterministic time evolution in a metastable state.
The hydrodynamic equations alone thus fail to provide a selection principle for the physically realized NESS.
Stochastic properties, as discussed in modern frameworks for NESS such as large deviation theory \cite{Touchette09,Bertini15} and stochastic thermodynamics \cite{Sekimoto-book, Seifert12}, 
are therefore essential for the selection of the unique steady state \cite{Global-modelB}.

In equilibrium, the selection problem is resolved by a thermodynamic variational principle, such as minimizing free energy, which uniquely determines the stable state \cite{Callen, Landau-Lifshitz-Statphys, Einstein}. 
This motivates extending such a variational approach to non-equilibrium systems.
As a variational principle in weak NESS, the minimum entropy production principle, originating from linear irreversible thermodynamics \cite{Onsager31, Groot-Mazur}, was proposed for the selection of a steady current \cite{Prigogine}. Despite the mathematically precise formulation of the minimum entropy production principle \cite{Maes-Netocny07,Maes-Netocny-SP13}, it is still challenging to predict the true steady state among multiple steady states.

To tackle state selection and non-trivial phenomena specifically in flux-driven phase coexistence, global thermodynamics was proposed by extending equilibrium variational principles \cite{Global-PRL, Global-JSP, Global-PRR}.
Its ability to predict non-trivial effects, such as the stabilization of metastable states, has been demonstrated numerically \cite{KNS} and analytically \cite{Global-modelB}.
This work extends global thermodynamics to incorporate gravity (mechanical driving) alongside heat flow (non-equilibrium driving).
Addressing this combination is crucial for developing a general thermodynamic description applicable to ubiquitous physical scenarios involving competing effects and inhomogeneities.
We achieve this by first constructing a consistent thermodynamic formulation for equilibrium systems under weak gravity \cite{Global-Glavity-eq}, which then serves as the basis for developing a variational function tailored for the NESS under the combined influences \cite{Global-Gravity-Heat}.

This extended formulation leads to a key insight: the effects of both gravity and heat flow are unified into a single parameter, the ``effective gravity'' ($g_\eff$), within our variational framework. 
This unification dictates the stable phase configuration, providing a powerful tool to predict and quantitatively describe the resulting NESS, which constitutes our central finding.

\begin{figure}[bt]
\begin{center}
\includegraphics[width=4.5cm]{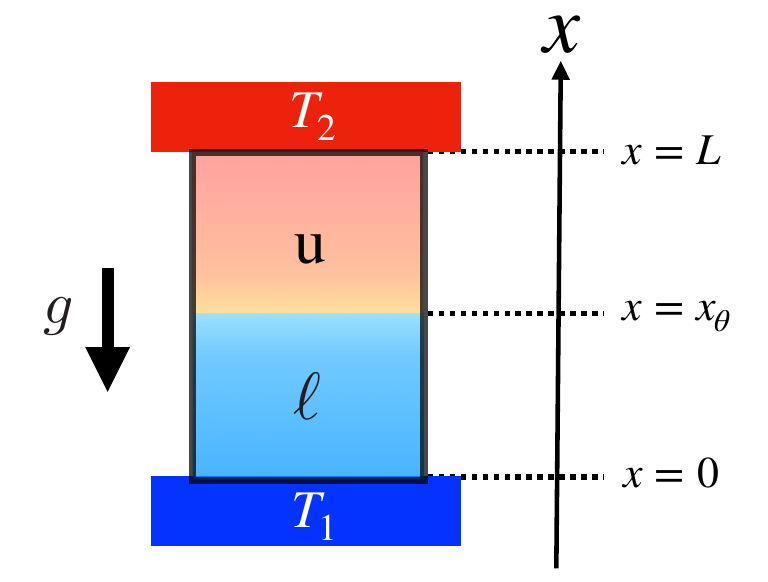}
\end{center}
\caption{System setup exposed to gravity and heat flow.}
\label{fig:setup}
\end{figure}

\paragraph*{System setup.---}
We consider $N$ particles of mass $m$ confined in a rectangular container of volume $V$, height $L$, and cross-sectional area $A=V/L$. See Fig.~\ref{fig:setup}. The vertical coordinate $x$ runs from the bottom wall ($x=\xb$) to the top wall ($x=\xt$). Uniform gravity $g$ acts in the negative $x$ direction, resulting in a gravitational potential energy $mg(x-\xo)$ relative to an arbitrary reference point $\xo$. The walls at $\xb$ and $\xt$ are maintained at constant temperatures $\Tb$ and $\Tt$, respectively. We define the temperature difference parameter $\Xi \equiv \Tt - \Tb$.
The total particle number $N$ and volume $V$ are fixed.

While the system is globally in a non-equilibrium steady state,
we utilize quantities based on local properties to construct a global thermodynamic description.
We define the global temperature $\bT$ as the density-weighted average
\begin{equation}
\bT \equiv \frac{A}{N} \int_{\xb}^{\xt} T(x) \rho(x) dx, \label{e:bT-def}
\end{equation}
where $\rho(x)$ is the number density profile. 
We also define a small dimensionless parameter $\ep$ representing the magnitude of the non-equilibrium and gravitational effects
\begin{equation}
\ep \equiv \max \left( \frac{|\Xi| }{\bT}, \frac{m g L}{\kB \bT}\right), \label{eq:epsilon_def}
\end{equation}
focusing on the linear response regime in $g$ and $\Xi$.
In the linear response regime, convection is absent, and the profiles (e.g., temperature) are linear within each single-phase region.

\paragraph*{Brief review of global thermodynamics of equilibrium systems with gravity.---}

Let $X$ denote the vertical position of the center of mass.
From statistical mechanics, the free energy of this system is expressed as
\begin{equation}
F_g(T,V,N,mg)=F(T,V,N)+Nmg (X- \xo)
\label{F-start}
\end{equation}
neglecting $O(\ep^2)$ terms.
To define $F_g $ uniquely, we impose the condition
\begin{equation}
F_g(T,V,N,-mg)=F_g(T,V,N,mg) ,
\label{symmetry}
\end{equation}
representing the reflection symmetry of thermodynamic properties with respect to $x$.
This condition leads to
\begin{align}
\xo=\xm\equiv\frac{\xb+\xt}{2}
\end{align}
as the unique reference point required by our global thermodynamic formulation for equilibrium systems under gravity \cite{Global-Glavity-eq}. 

Using statistical mechanics, $F_g$ can be shown to satisfy the differential relation
\begin{align}
dF_g=-SdT-\bar p dV+\bar\mu dN-\Psi d(mgL) +O(\ep^2),
\label{Fg-relation-0}
\end{align}
where $\bar p$ and $\bar \mu$ are the spatially averaged pressure
and chemical potential over the whole system. $S$ is the thermodynamic
entropy, and $\Psi$ is defined as
\begin{align}
\Psi\equiv N\frac{\xm- X}{L} .
\label{psi-def}
\end{align}
The expression \eqref{Fg-relation-0} implies that the thermodynamic state space $(T,V,N)$ is extended to $(T,V,N,mgL)$.
$F_g$ is now expressed as
\begin{equation}
F_g(T,V,N,mgL)=F(T,V,N)-\Psi mgL
\label{e:Fg-eq-thermo0}
\end{equation}
with the fundamental relation \eqref{Fg-relation-0} of thermodynamics.

$\Psi$ is non-zero only during liquid-gas coexistence. In this case, $\Psi$ is discontinuous at $g=0$ because $X<\xm$ for $g >0$ (assuming liquid is denser) and $X>\xm$ for $g <0$.
To determine the equilibrium configuration in liquid-gas coexistence, we construct a variational function $\mathcal{F}_g$, encompassing the equilibrium free energy $F_g$.
Let $\subl$ and $\subu$ denote the lower ($\xb<x<\xs$) and upper ($\xs<x<\xt$) phases, respectively, which can be either liquid or gas.
Here, $\xs$ is the interface position.
Then, let $\Vl$ and $\Nl$ be the volume and particle number of the lower phase.
From geometrical considerations, \eqref{psi-def} can be expressed as
\begin{align}
\Psi(\Nl, \Vl)=\frac{N}{2}\left(\frac{\Nl}{N}-\frac{\Vl}{V} \right).
\label{e:Psi-NlVl}
\end{align}
Referring to \eqref{e:Fg-eq-thermo0} and considering the additivity of free energy, the variational function used to determine the equilibrium value of $(\Vl, \Nl)$ is given by 
\begin{align}
&\mathcal{F}_g(\Nl,\Vl ; T,V,N,mgL)\equiv F(T,\Vl,\Nl)\nm
&~~~~+F(T,V-\Vl,N-\Nl)-\Psi(\Nl, \Vl) mgL. 
\label{e:var-eq}
\end{align}
This represents the free energy assuming phase separation under gravity.

The principle of minimum free energy implies that
\begin{align}
F_g(T,V,N, mgL)=\min_{\Vl, \Nl}\mathcal{F}_g(\Nl, \Vl ; T,V,N,mgL).
\end{align}
Minimizing $\mathcal{F}_g$ with respect to $(\Nl,\Vl)$ yields the equilibrium values $(\NleL,\VleL)$, corresponding to the configuration where the liquid phase occupies the lower region (assuming liquid is denser).
The indices $\subL$ and $\subG$ specify liquid and gas, respectively.

Let $\NL_0$ and $\VL_0$ be the liquid amounts at $g=0$ determined by the lever rule.
The saturated specific volumes $v_\subC^\subL(T)$ and $v_\subC^\subG(T)$ for liquid and gas are given by
$v_\subC^\subL(T)=\VL_0/\NL_0$ and $v_\subC^\subG(T)=\VG_0/\Ng_0$ with $\VG_0=V-\VL_0$ and $\Ng_0=N-\NL_0$.
We then have $\NleL=\NL_0+O(\ep)$ and $\VleL=\VL_0+O(\ep)$.
Combining this result with the force balance equation yields the standard equilibrium phase coexistence conditions: continuity of local pressure and local chemical potential across the interface.

\paragraph*{Global thermodynamics of heat conduction systems with gravity.---}

We now extend the global thermodynamic framework to describe heat conduction systems under gravity.
We assume that the variational function for the NESS takes the same form as \eqref{e:var-eq}, but with the equilibrium temperature $T$ replaced by appropriate global temperatures for each phase.
Specifically, we propose the variational function:
\begin{align}
&\mathcal{F}_g(\Nl,\Vl ; \bT,V,N,mgL,\Xi)\equiv 
F^\subl(\bT^\subl,\Vl,\Nl)\nm
&~~~~ +F^\subu(\bT^\subu,V-\Vl,N-\Nl)
-\Psi(\Nl, \Vl) mgL, 
\label{e:var-ss}
\end{align}
with $\Psi(\Nl,\Vl)$ given by \eqref{e:Psi-NlVl}.
Here, $F^\subl$ and $F^\subu$ represent the free energies associated with the lower ($\subl$) and upper ($\subu$) phases depending on the global temperatures $\bT^\subl$ and $\bT^\subu$ of the respective regions
\begin{align}
&F^\subl(\bT^\subl,\Vl,\Nl)=A\int_{\xb}^{\xs} f(T(x), \rho(x)) dx,\nm
&F^\subu(\bT^\subu,V-V^\subl,N-N^\subl)=A\int_{\xs}^{\xt} f(T(x), \rho(x))dx,\nonumber
\end{align}
with the local free energy density $f(T(x), \rho(x))$
allowing for locally metastable states.
By definition, the global temperatures satisfy
$N\bT=\Nl\bT^\subl+(N-\Nl)\bT^\subu$.
Assuming approximately linear temperature profiles within each phase, we have $\bT^\subl=(\Tb+\theta)/2$ and $\bT^\subu=(\Tt+\theta)/2$, where $\theta$ is the interface temperature. 
Combining these relations, we obtain
\begin{align}
\bT^\subl=\bT-\frac{\Xi}{2}\frac{N-\Nl}{N}, \quad
\bT^\subu=\bT+\frac{\Xi}{2}\frac{\Nl}{N}.
\label{e:bT-lu}
\end{align}

The NESS is assumed to minimize the variational function.
Potential steady state configurations correspond to the stationary points of $\mathcal{F}_g$, found by solving the variational equations
\begin{align}
\pder{\mathcal{F}_g}{\Vl}=0, \qquad 
\pder{\mathcal{F}_g}{\Nl}=0, 
\label{e:var-neq}
\end{align}
where the differentiations are taken at fixed $(\bT,V,N,mgL,\Xi)$.
These two equations (see Appendix \ref{a:mu-balance} for details) result in pressure continuity at the interface and a non-trivial chemical potential balance condition
\begin{align}
&\mu^\subl(T_\subS,p_\theta)=\mu^\subu(T_\subS,p_\theta),
\label{e:mu-continuous-neq}\\
&T_\subS\equiv \bT+\frac{\Xi}{2}\frac{2\Nl-N}{N},
\label{e:Ts}
\end{align}
where $p_\theta$ is the pressure at the interface.
The relation \eqref{e:mu-continuous-neq} leads to
\begin{align}
T_\subS=\Tc(p_\theta)
\label{e:Tc}
\end{align}
with the equilibrium transition temperature $\Tc(p_\theta)$ at $p_\theta$.

Remarkably, \eqref{e:mu-continuous-neq}, \eqref{e:Ts}, and \eqref{e:Tc} are consistent with the results reported for the zero-gravity ($g=0$) heat conduction case in \cite{Global-PRL,Global-JSP}. Specifically, they imply that the interface temperature $\theta$ generally deviates from the equilibrium coexistence temperature $\Tc(p_\theta)$,
\begin{align}
\theta=\Tc(p_{\theta})+J\left(\frac{\bar v}{v_\subC^\subl\kappa^\subu}-\frac{\bar v}{v_\subC^\subu\kappa^\subl}\right)\frac{\xs(L-\xs)}{2L}+O(\ep^2),
\label{e:theta-J}
\end{align}
as a physical effect of the linear response regime $O(\ep)$.
Here, $J$ is the heat flux (defined positive from bottom to top), $\kappa^{\subl,\subu}$ are the thermal conductivities of the phases, and $\bar{v}=V/N$. 
This consistency suggests that the stabilization of metastable states by heat flux is a robust phenomenon, persisting even under gravity.
This non-trivial prediction has been numerically confirmed for order-disorder coexistence in the Hamiltonian Potts model \cite{KNS} and analytically verified for liquid-gas coexistence in boundary-driven diffusive systems \cite{Global-modelB}.

\paragraph*{Effective gravity induced by heat flux.---}

\begin{figure}[bt]
\begin{center}
\includegraphics[width=6.5cm]{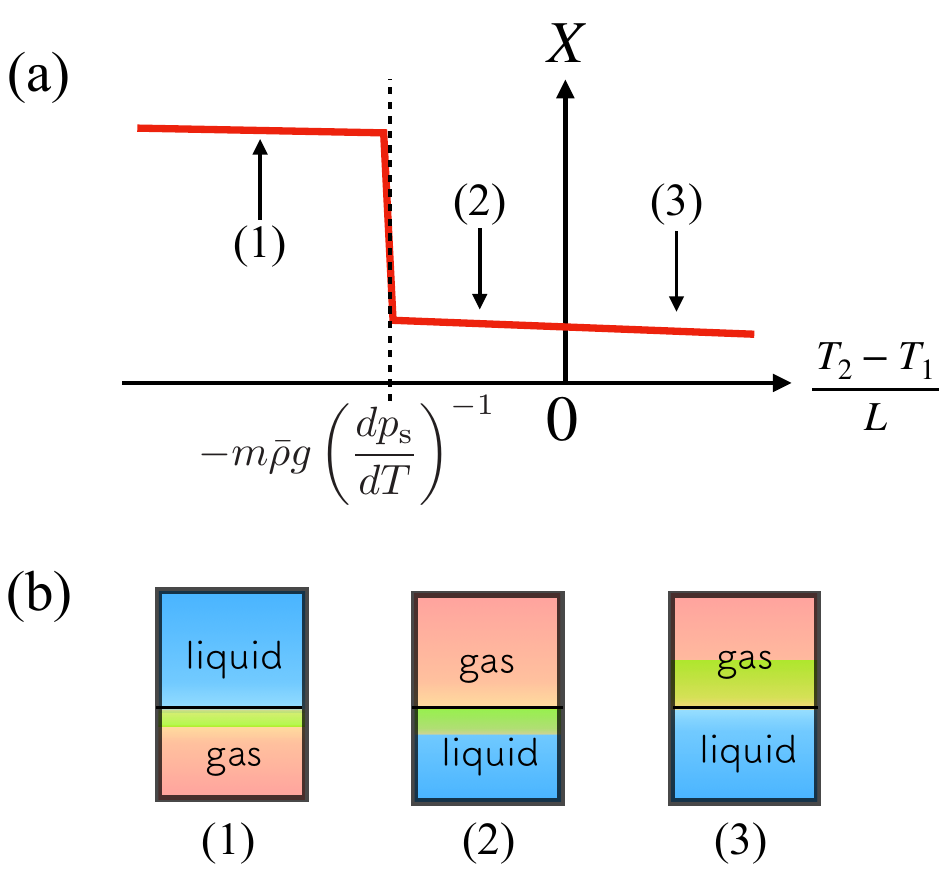}
\end{center}
\caption{(a) Center of mass $X$ as a function of $(T_2-T_1)/L$ under fixed gravity. $X$ shows a discontinuous transition when $g_\eff$ changes sign. 
(b) Configuration of the system for each point indicated in (a). The light green regions are metastable: supercooled gas in (1) and (3) and superheated liquid in (2).} 
\label{fig:config}
\end{figure}

Note that the stationarity conditions derived from \eqref{e:var-neq} generally admit two solutions: one corresponding to liquid below gas, denoted $(\NlsL, \VlsL)$, and the other to gas below liquid, denoted $(\NlsG, \VlsG)$.
The NESS corresponds to the configuration that yields the global minimum of the variational function $\mathcal{F}_g$ defined in  \eqref{e:var-ss}.

Substituting \eqref{e:bT-lu} into \eqref{e:var-ss}, expanding in $\ep$, and neglecting $O(\ep^2)$ terms (see Appendix \ref{a:geff-derivation}), we obtain
\begin{align}
&\mathcal{F}_g(\NlsL,\VlsL) = F(\bT,V,N)-\Psi(\NL_0,\VL_0)mg_\eff L, \label{e:Fg-LG}\\ 
&\mathcal{F}_g(\NlsG,\VlsG) = F(\bT,V,N)+\Psi(\NL_0,\VL_0)mg_\eff L, \label{e:Fg-GL}
\end{align}
where $g_\eff$ is the effective gravitational acceleration,
\begin{align}
g_\eff\equiv g+\frac{\bar{v}}{m}\frac{d\Ps}{dT}\bigg|_{\bT} \frac{\Xi}{L}
\label{e:g-eff}
\end{align}
with the saturation pressure $\Ps(T)$ at temperature $T$.
In \eqref{e:Fg-GL}, $\Psi(N^\subG_0,V^\subG_0)=-\Psi(\NL_0,\VL_0)$ was used, where $\NL_0$ and $\VL_0$ are the zero-gravity equilibrium amounts.

The difference between \eqref{e:Fg-LG} and \eqref{e:Fg-GL}, 
\begin{align}
\mathcal{F}_g(\NlsG,\VlsG)-\mathcal{F}_g(\NlsL,\VlsL)
= 2mg_\eff L\Psi(\NL_0,\VL_0),
\label{e:Fg-diff-twomin-neq}
\end{align}
is generally non-zero. 
The lever rule yields the expressions of $\NL_0$ and $\VL_0$ with the saturated specific volumes.  Using them, \eqref{e:Psi-NlVl} gives
\begin{align}
\Psi(\NL_0,\VL_0)=\frac{N}{2}\frac{(v^\subG_\subC(\bT)-\bar{v})(\bar{v}-v^\subL_\subC(\bT))}{\bar{v}(v^\subG_\subC(\bT)-v^\subL_\subC(\bT))}.
\label{e:Psig-LG}
\end{align}
For liquid-gas coexistence, we assume $v_\subC^\subL(\bT)<\bar{v}<v_\subC^\subG(\bT)$, which ensures $\Psi(\NL_0,\VL_0)>0$.
We conclude that
\begin{align}
&\mathcal{F}_g(\NlsL,\VlsL)<\mathcal{F}_g(\NlsG,\VlsG) \qquad (g_\eff >0), \label{e:minimaLG}\\
&\mathcal{F}_g(\NlsG,\VlsG)<\mathcal{F}_g(\NlsL,\VlsL) \qquad (g_\eff <0).
\label{e:minimaGL}
\end{align}

The configuration with the smaller value of $\mathcal{F}_g$ corresponds to the stable NESS.
The sign of $g_\eff$ dictates the configuration: liquid is at the bottom if $g_\eff > 0$ and at the top if $g_\eff < 0$, directly answering the question posed in the Introduction.
This can be explicitly expressed by
\begin{align}
X=\xm-L\frac{|g_\eff|}{g_\eff} \frac{\Psi(N^\subL_0,V^\subL_0)}{N}+O(\ep)
\end{align}
with \eqref{e:Psig-LG}.
Figure \ref{fig:config}(a) demonstrates how the heat flux alters the steady state under a fixed gravity.
This stabilization occurs when the entropic contribution to the variational function $\mathcal{F}_g$, quantified by the thermal term in $g_{\mathrm{eff}}$, overcomes the increase in gravitational potential energy.

When the top is cooled sufficiently such that the mean temperature gradient $\Xi/L$ satisfies the condition $g_\eff < 0$, the liquid floats above the gas.
The transition point $g_\eff=0$ corresponds to the condition 
\begin{align}
\Ps(\Tb)-\Ps(\Tt)=p_1-p_2,
\label{e:criteria}
\end{align}
where $p_1$ and $p_2$ are the pressures at the bottom and the top. 
This follows from \eqref{e:g-eff}  with the relations
$(d\Ps/dT)\Xi=\Ps(T_2)-\Ps(T_1)+O(\ep^2)$ and $p_1-p_2 = mg L/\bar{v}$. 
The criterion \eqref{e:criteria} explicitly expresses the balance between thermal driving (related to saturation pressure difference) and mechanical driving (gravity).
See Appendix \ref{a:Exp-setup} for an example of experimental setup.

When $g>0$ and $\Xi>0$ ($\Tt>\Tb$), gravity and heat flux are aligned, and the gas settles at the top as expected. However, unexpectedly, a significant portion of the gas can become supercooled by the heat flux as shown in \eqref{e:theta-J} [Fig.~\ref{fig:config}(b)-(3)].
If $\Tt$ is reduced slightly below $\Tb$ while $g_\eff$ remains positive, the ``normal'' configuration (liquid at bottom) persists.
If $\Xi<0$ ($\Tb>\Tt$) with $g_\eff > 0$, the liquid-at-bottom configuration is unchanged, but phase states are altered: the gas becomes stable, and the liquid near the interface superheats [Fig.~\ref{fig:config}(b)-(2)].
This superheated liquid, absent at $g=0$, is thus a gravity-induced non-equilibrium effect.
When $\Tt$ is low enough for $g_\eff<0$, the liquid floats above the gas. Here, the liquid is stable, while the gas near the interface is supercooled [Fig.~\ref{fig:config}(b)-(1)].
While configurations (1) and (3) may appear as spatial inversions, the intermediate state (2) with superheated liquid implies a more complex interplay. Further investigation is warranted \cite{Global-Gravity-Heat}.

\begin{figure}[bt]
\begin{center}
\includegraphics[width=7.cm]{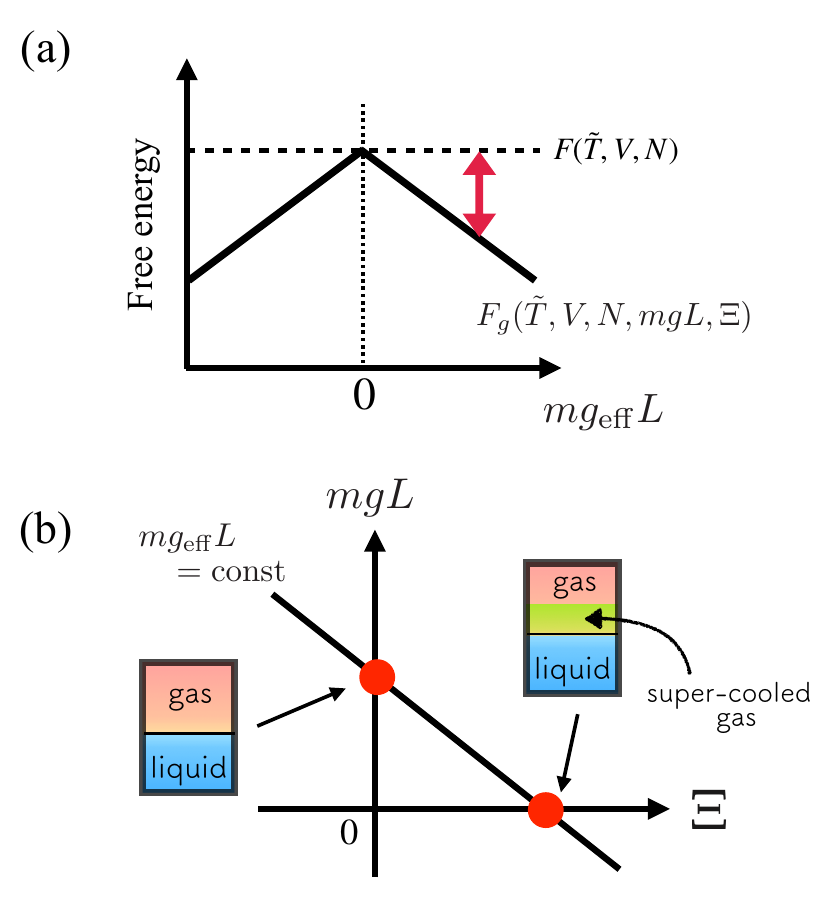}
\end{center}
\caption{(a) Thermodynamic free energy $F_g$ as a function of $mg_\eff L$ at fixed $(\bT,V,N)$. (b) Correspondence between gravity and temperature gradient. The value of $F_g$ is kept constant along the solid line $mg_\eff L=\mathrm{const}$, but the local state inside the system differs.} 
\label{fig:Fg}
\end{figure}

\paragraph*{Non-analyticity of free energy.--} 

The thermodynamic free energy for the NESS is given by
\begin{align}
F_g(\bT,V,N, mgL,\Xi)=\min_{\Vl, \Nl}\mathcal{F}_g(\Vl, \Nl ; \bT,V,N,mgL,\Xi).
\end{align}
Considering \eqref{e:minimaLG} and \eqref{e:minimaGL}, this minimization results in
\begin{align}
F_g(T,V,N,mgL,\Xi)=F(T,V,N)-\Psi(\NL_0,\VL_0) m|g_\eff|L,
\label{e:Fg-neq-thermo}
\end{align}
which exhibits a non-analyticity (a cusp) at $g_\eff=0$ as shown in Fig.~\ref{fig:Fg}(a).
It signifies the first-order transition between the two configurations, whether the denser phase settles lower or upper. 
We emphasize that this transition also occurs in equilibrium when $g$ changes sign, reflecting the spatial inversion symmetry of the system.

The NESS free energy \eqref{e:Fg-neq-thermo} indicates
\begin{align}
F_g(\bT,V,N,mgL,\Xi)=F_g^{\eq}(\bT,V,N,mg_\eff L),
\end{align}
where $F_g^{\eq}$ is the equilibrium free energy function. 
The NESS free energy retains the functional form of the equilibrium one, but with $g$ replaced by $g_\eff$.
Figure \ref{fig:Fg}(b) illustrates a line of constant free energy in the $(\Xi, mgL)$ plane for fixed $\bT$, $V$, and $N$. 
An equilibrium state ($\Xi=0$) can be connected to a pure heat conduction state ($g=0$) along a line of constant $g_\eff$ without changing the value of $F_g$.
As indicated by \eqref{e:theta-J} and illustrated in Fig.~\ref{fig:Fg}(b), the interface temperature $\theta$ deviates from $\Tc(p_\theta)$ along this line, leading, for example, to the stabilization of supercooled gas.

\paragraph*{Concluding remarks. --}

We have extended global thermodynamics to unify gravity and heat flow, yielding a variational principle governed by an effective gravity, $g_\eff$, that predicts the stable phase configuration. 
This work also confirms the stabilization of local metastable states, a prediction of earlier global thermodynamics studies. 
Deriving our thermodynamic potential from microscopic models remains an important future challenge.

Our predictions could be tested by molecular dynamics (MD) simulations. 
Indeed, the liquid floating phenomenon has been observed numerically \cite{ANS,Wagner}, in qualitative agreement with our theory. 
While these studies assumed local equilibrium at the interface, their slow convergence towards the thermodynamic limit suggests that deviations from local equilibrium, as predicted by our principle, may play a crucial role.

Experimental verification is crucial but challenging. 
Quantitative reports are lacking, primarily due to unwanted convection, which can be induced by horizontal temperature differences as small as $1$~K \cite{Higa-Kagawa}. 
Precise experiments to test global thermodynamics would require careful control, such as active control \cite{Howle97}, and should focus on the steady interface properties or the configurational transition at $g_\eff=0$.

Finally, we comment on Luttinger's theory \cite{Luttinger}, which may seem to resonate conceptually with the concept of $g_\eff$. In Luttinger's formalism, heat conduction is modeled using an external potential, interpreted as a gravitational effect, in the Hamiltonian.
Setting aside the validity of this physical interpretation, we view Luttinger's theory as a mechanical model that yields heat conduction.
In contrast, our $g_\eff$ emerges from a thermodynamic variational principle for NESS selection.
Investigating the thermodynamic potential for a heat conduction system under gravity within the Luttinger framework could be an interesting direction for future research, and might reveal a different type of effective gravity or potentially converge with the one proposed in this Letter.

\paragraph*{Acknowledgement.--}

The authors thank A. Yoshida, F. Kagawa, R. Higa, and A. Hisada for useful discussions.
This study was supported by JSPS KAKENHI Grant Numbers JP23K22415, JP25K00923, JP25K22002, and JP25H01975.


\twocolumngrid
\appendix
\normalsize

\renewcommand{\theequation}{\Alph{section}\arabic{equation}}
\renewcommand{\thefigure}{\Alph{section}\arabic{figure}}
\renewcommand{\thetable}{\Alph{section}\arabic{table}}

\setcounter{equation}{0}
\setcounter{figure}{0}
\setcounter{table}{0}

\vspace{1cm}

\begin{center}
{\large \bf End Matter}
\end{center}

\section{Derivation of Eq.~(\ref{e:mu-continuous-neq})}
\label{a:mu-balance}

red{
In the linear response regime, we assume that local quantities exhibit linear profiles within each phase. Therefore, each state  can be represented by its mean values. This allows us to map them to effective equilibrium systems, $(\bT^\subl, \mathcal{V}^\subl, \mathcal{N}^\subl)$ and $(\bT^\subu, \mathcal{V}^\subu, \mathcal{N}^\subu)$, for which the following thermodynamic relations hold,
\begin{align}
&dF^\subl=-S^\subl d\bT^\subl-\bP^\subl d\mathcal{V}^\subl+\mu^\subl d\mathcal{N}^\subl,\\
&dF^\subu=-S^\subu d\bT^\subu-\bP^\subu d\mathcal{V}^\subu+\mu^\subu d\mathcal{N}^\subu,
\end{align}
where $\bP^\subl$ and $\bP^\subu$ are the mean pressures, and $\mu^\subl=\mu^\subl(\bT^\subl,\bP^\subl)$ and $\mu^\subu=\mu^\subu(\bT^\subu,\bP^\subu)$ are the chemical potentials. The entropies are given by $S^\subl=\mathcal{N}^\subl s_\subC^\subl(\bT^\subl)$ and $S^\subu=\mathcal{N}^\subu s_\subC^\subu(\bT^\subu)$, where $s_\subC$ is the specific entropy.
}

Because $\bT^\subl$ and $\bT^\subu$ are independent of $\mathcal{V}^\subl$ (see \eqref{e:bT-lu}), substituting the derivatives of the variational function \eqref{e:var-ss} into the first equation of \eqref{e:var-neq} yields
\begin{align}
\bP^\subl-\bP^\subu=\frac{NmgL}{2V}
\label{eq:app_force_balance_mean}
\end{align}
using the relation $\partial \Psi/\partial{\mathcal{V}^\ell}=-N/(2V)$ from \eqref{e:Psi-NlVl}.
This relation \eqref{eq:app_force_balance_mean} reflects the overall force balance $(p_1 - p_2)A = Nmg$ and implies pressure continuity at the interface.

Similarly, the second equation of \eqref{e:var-neq} yields
\begin{align}
&\mu^\subl-\mu^\subu-(S^\subl+S^\subu)\frac{\Xi}{2N}= \frac{mgL}{2}. \label{e:varN-neq-1_app}
\end{align}
Using $T_\subS$ defined in \eqref{e:Ts} of the main text, the global temperatures for each phase, given by \eqref{e:bT-lu}, can be written as
\begin{align}
\bT^\subl = T_\subS - \frac{\Xi}{2}\frac{\Nl}{N}, \quad
\bT^\subu = T_\subS + \frac{\Xi}{2}\frac{N-\Nl}{N}.
\label{eq:app_Tb_rel_Ts_revised}
\end{align}
Furthermore, the mean pressures $\bP^\subl$ and $\bP^\subu$ relate to the interface pressure $p_\theta$ via the force balance relations for each phase:
\begin{align}
(\bP^\subl-p_\theta)A=\frac{\Nl mg}{2},\quad
(\bP^\subu-p_\theta)A=-\frac{(N-\Nl) mg }{2}.
\label{eq:app_phase_force_balance_revised}
\end{align}
Expanding $\mu^\subl(\bT^\subl,\bP^\subl)$ and $\mu^\subu(\bT^\subu,\bP^\subu)$ around $(T_\subS,p_\theta)$ up to $O(\ep)$, using $\partial\mu/\partial T = -S/N$, $\partial\mu/\partial p = V/N$, $\Vl=A\xs$, and $V-\Vl=A(L-\xs)$ along with \eqref{eq:app_Tb_rel_Ts_revised} and \eqref{eq:app_phase_force_balance_revised}, we have
\begin{align}
&\mu^\subl(\bT^\subl,\bP^\subl) = \mu^\subl(T_\subS,p_\theta) + \frac{\Xi}{2}\frac{S^\subl}{N} + \frac{mg \xs}{2} + O(\ep^2), \label{eq:app_mu_l_exp_revised} \\
&\mu^\subu(\bT^\subu,\bP^\subu) = \mu^\subu(T_\subS,p_\theta) - \frac{\Xi}{2}\frac{S^\subu}{N} - \frac{mg (L-\xs)}{2} + O(\ep^2). \label{eq:app_mu_u_exp_revised}
\end{align}
Substituting \eqref{eq:app_mu_l_exp_revised} and \eqref{eq:app_mu_u_exp_revised} into \eqref{e:varN-neq-1_app}, and neglecting terms of $O(\ep^2)$, we obtain \eqref{e:mu-continuous-neq} of the main text: $\mu^\subl(T_\subS,p_\theta)=\mu^\subu(T_\subS,p_\theta)$.

\section{Derivation of Eqs.~(\ref{e:Fg-LG}) and (\ref{e:Fg-GL})} 
\label{a:geff-derivation}

Consider the solution $(\NlsL,\VlsL)$ where the liquid phase settles in the lower region. 
We write
\begin{align}
\NlsL=\NL_0+\Delta N, \quad \VlsL=\VL_0+\Delta V,
\end{align}
where $\NL_0$ and $\VL_0$ are the particle number and volume of the liquid phase at $g=0$ and $\Xi=0$ (determined by the lever rule), and $\Delta N, \Delta V$ are deviations of $O(\ep)$.
According to \eqref{e:var-ss} in the main text, the variational function is
\begin{align}
&\mathcal{F}_g(\NlsL,\VlsL) =F^\subl(\bT^\subl,\VlsL,\NlsL)\nm
&~+F^\subu(\bT^\subu,V-\VlsL,N-\NlsL) - \Psi(\NlsL, \VlsL) mgL. 
\label{e:F-apx0_revised} 
\end{align}
To leading order in $\ep$, we use the approximations:
\begin{align}
&\Psi(\NlsL, \VlsL) 
= \frac{N}{2}\left(\frac{\NL_0}{N}-\frac{\VL_0}{V} \right) + O(\ep), \label{eq:Psi_approx_app} \\
&\bT^\subl=\bT-\frac{\Xi}{2}\frac{\Ng_0}{N} + O(\ep^2), \label{eq:Tb_l_approx_app} \\ 
&\bT^\subu=\bT+\frac{\Xi}{2}\frac{\NL_0}{N} + O(\ep^2). \label{eq:Tb_u_approx_app}
\end{align}
Expanding the sum of the first two free energy terms in \eqref{e:F-apx0_revised} with respect to $\ep$ leads to
\begin{align}
&F^\subl(\bT^\subl,\VlsL,\NlsL)+F^\subu(\bT^\subu,V-\VlsL,N-\NlsL) \nm
&= F(\bT,\VL_0,\NL_0)+F(\bT,\VG_0,\Ng_0) \nm 
&\quad - S_0^\subL(\bT^\subl-\bT) - p_0^\subL \Delta V + \mu_0^\subL \Delta N \nm 
&\quad - S_0^\subG(\bT^\subu-\bT) - p_0^\subG (-\Delta V) + \mu_0^\subG (-\Delta N) + O(\ep^2).
\label{e:F-expand0}
\end{align} 
Here, the subscript `0' indicates the quantity at zero-gravity in equilibrium. 
Using the equilibrium conditions at $O(\ep^0)$, $p_0^\subL=p_0^\subG$ and $\mu_0^\subL=\mu_0^\subG$, the terms proportional to $\Delta V$ and $\Delta N$ cancel out. 
Using specific entropies $s_\subC^\subL(\bT)$  and $s_\subC^\subG(\bT)$, 
$S_0^\subL=s_\subC^\subL(\bT) \NL_0$ and $S_0^\subG=s_\subC^\subG(\bT) \Ng_0$.
The sum $F(\bT,\VL_0,\NL_0)+F(\bT,\VG_0,\Ng_0)$ equals $F(\bT,V,N)$.
Applying these and substituting \eqref{eq:Tb_l_approx_app} and \eqref{eq:Tb_u_approx_app},
\eqref{e:F-expand0} is further transformed as
\begin{align}
& F(\bT,V,N) 
+ \frac{\Xi}{2N} \NL_0 \Ng_0 (s_\subC^\subL - s_\subC^\subG).
\label{e:F-apx1_revised} 
\end{align}
Applying the Clausius-Clapeyron relation, $s_\subC^\subL - s_\subC^\subG = (dp_s/dT)_{\bT}(v_\subC^\subL - v_\subC^\subG)$ with $v_\subC^\subL=\VL_0/\NL_0$ and $v_\subC^\subG=\VG_0/\Ng_0$, the second term becomes
\begin{align}
 \frac{\Xi}{2N} \der{\Ps}{T}\bigg|_{\bT} (\VL_0 \Ng_0 - \VG_0 \NL_0).
\label{e:trans1}
\end{align}
Using $\VL_0 \Ng_0 -\VG_0 \NL_0 = -(V\NL_0 - N\VL_0)$ and $\Psi(\NL_0,\VL_0) = \frac{1}{2V} (\NL_0 V - N \VL_0)$, \eqref{e:trans1} is further transformed into
\begin{align}
-\Xi\der{\Ps}{T}\bigg|_{\bT}\frac{V}{N}\Psi(\NL_0,\VL_0).
\end{align}
Substituting this result and \eqref{eq:Psi_approx_app} into \eqref{e:F-apx0_revised}, and neglecting $O(\ep^2)$ terms, we obtain
\begin{align}
&\mathcal{F}_g(\NlsL,\VlsL) =F(\bT,V,N)\nm
&\qquad -\Psi(\NL_0,\VL_0) mgL -\Xi\der{\Ps}{T}\bigg|_{\bT}\frac{V}{N}\Psi(\NL_0,\VL_0) + O(\ep^2) \nm
&=F(\bT,V,N)-\Psi(\NL_0,\VL_0) \left( mg L+\Xi\der{\Ps}{T}\bigg|_{\bT}\frac{V}{N}\right) + O(\ep^2),\nonumber
\end{align}
which is \eqref{e:Fg-LG} in the main text (upon identifying $V/N = 1/\bar\rho$ and neglecting $O(\ep^2)$).

A similar procedure for the other stationary solution $(\NlsG,\VlsG)$ (gas phase in the lower region), or by noting the symmetry $\Psi(\Ng_0,\VG_0)=-\Psi(\NL_0,\VL_0)$ (where $\Ng_0 = N-\NL_0$, $\VG_0 = V-\VL_0$), yields \eqref{e:Fg-GL}:
\begin{align}
&\mathcal{F}_g(\NlsG,\VlsG) \nm
&=F(\bT,V,N)-\Psi(\Ng_0,\VG_0) \left( mg L+\Xi\der{\Ps}{T}\bigg|_{\bT}\frac{V}{N}\right) + O(\ep^2) \nm
&=F(\bT,V,N)+\Psi(\NL_0,\VL_0) \left( mg L+\Xi\der{\Ps}{T}\bigg|_{\bT}\frac{V}{N}\right) + O(\ep^2).\nonumber
\end{align}

\section{An Example of Experimental Setup} 
\label{a:Exp-setup}

Below, we consider an experimental setup involving a cylindrical container of radius $r$ with thermally insulated side walls. 
The bottom and top walls are maintained at constant temperatures $\Tb$ and $\Tt$, respectively.
For the working fluid, in this example \ce{H2O}, its saturation pressure $\Ps(T)$ as a function of temperature can be obtained from a standard thermodynamic database \cite{NIST}. 
The critical radius $r_c$ at which the transition occurs between the configuration with liquid below gas and that with liquid floating above gas is identified from the condition $g_\eff=0$. This leads to the expression 
\begin{align}
r_\subC(\Tb,\Tt)=\sqrt{\frac{Nmg}{\pi [\Ps(\Tb)-\Ps(\Tt)]}}
\label{e:rc} 
\end{align}
derived from the criterion given in \eqref{e:criteria} of the main text.
The liquid phase is expected to settle at the bottom if the container radius $r < r_\subC$, while it should float above the gas phase if $r > r_\subC$.
It is important to note that this expression for $r_\subC$ is independent of the total height $L$ of the container.

To ensure that the system is dominated by steady heat conduction, thereby allowing for a clear observation of the predicted phase inversion, convective instabilities within the fluid layers must be suppressed. For the liquid layer, this typically requires its Rayleigh number, $R_a$, to be less than the critical value for the onset of Rayleigh-B\'enard convection (approximately $1700$ for a layer between rigid boundaries). This requirement, combined with the conditions for maintaining liquid-gas coexistence and achieving a specific liquid volume fraction, imposes constraints on the permissible total height $L$ and the applied temperature difference $|\Tb-\Tt|$.

As a specific example, we consider a system containing 0.5 mol of \ce{H2O}. For a bottom temperature $\Tb=300 \text{ K}$ and a top temperature $\Tt=299.25 \text{ K}$, the critical radius from \eqref{e:rc}  is $r_\subC \approx 1.36 \text{ cm}$. Assuming a liquid volume fraction of 0.2, a consistent container height is $L \approx 7.82 \text{ cm}$. This results in a liquid layer height of $1.56 \text{ cm}$ and a Rayleigh number $R_a \approx 434$. Since $R_a \ll 1700$, convection is suppressed, providing a viable experimental condition to test the phase inversion around $r_c$.

\begin{figure}[b]
    \centering 
    \includegraphics[width=6.5cm]{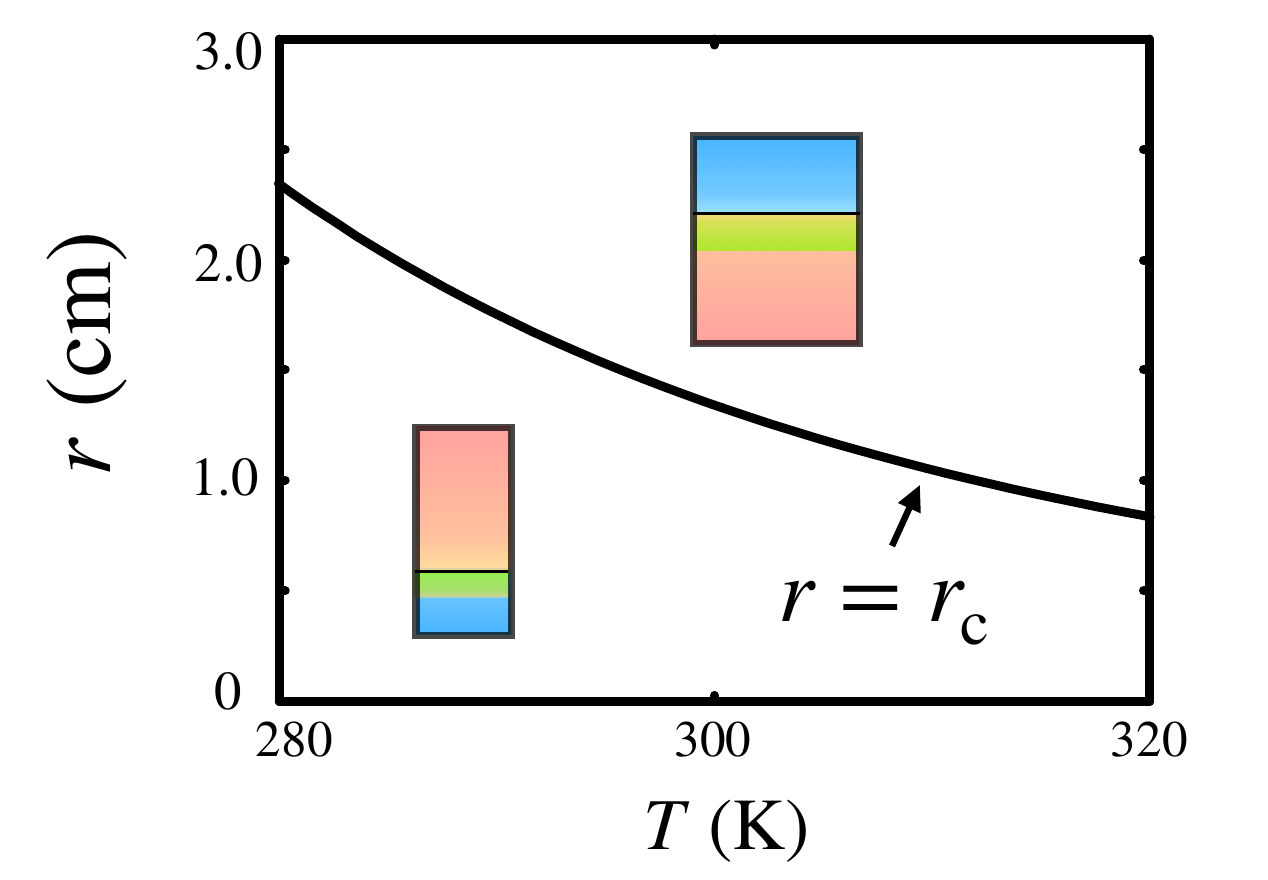}
    \caption{Critical radius $r_c$ for the configuration inversion of 0.5 mol of \ce{H2O} as a function of the average temperature $T \equiv (\Tb+\Tt)/2$. The temperature difference is held constant at $\Tb - \Tt = 0.75$ K. The line separates the liquid-below-gas ($r < r_\subC$) and gas-below-liquid ($r > r_\subC$) regimes.} 
    \label{fig:rc-T}
\end{figure}

\vfill

\end{document}